# Coding Improves the Throughput-Delay Trade-off in Mobile Wireless Networks

Zhenning Kong, *Student Member, IEEE,* Edmund M. Yeh, *Member, IEEE,*
Emina Soljanin, *Member, IEEE*

**Abstract**

We study the throughput-delay performance tradeoff in large-scale wireless ad hoc networks. It has been shown that the per source-destination pair throughput can be improved from $\Theta(1/\sqrt{n \log n})$ to $\Theta(1)$ if nodes are allowed to move and a 2-hop relay scheme is employed. The price paid for such an improvement on throughput is large delay. Indeed, the delay scaling of the 2-hop relay scheme is $\Theta(n \log n)$ under the random walk mobility model. In this paper, we employ coding techniques to improve the throughput-delay trade-off for mobile wireless networks. For the random walk mobility model, we improve the delay from $\Theta(n \log n)$ to $\Theta(n)$ by employing Reed-Solomon codes. Our approach maintains the diversity gained by mobility while decreasing the delay.

## I. INTRODUCTION

In large-scale wireless networks, a fundamental question is how much information one source node can transmit to a destination in the presence of other simultaneous transmissions when the number of nodes grows large. In their seminal paper [1], Gupta and Kumar investigate the network layer throughput, or the capacity, of large-scale wireless networks. They study the scaling behavior of the throughput in static random wireless networks, where the network has $n$ stationary nodes distributed in a unit disk. Every node serves simultaneously as a source for a randomly chosen destination, as a destination for another source, as well as a relay for all other source-destination (S-D) pairs. The authors show in [1] that when the nodes are randomly distributed, the per S-D pair throughput scales as $\Theta(1/\sqrt{n \log n})$,[1] and when the nodes are optimally placed, the throughput scales as $\Theta(1/\sqrt{n})$. This is the best achievable throughput performance even allowing for optimal scheduling, routing and relaying schemes. This result is somewhat pessimistic since as the network size grows, the throughput goes to zero.

In [2], Grossglausser and Tse propose a 2-hop relay scheme and show that a surprisingly significant improvement on throughput can be achieved if nodes are allowed to move. Indeed, they show that when

This research is supported in part by National Science Foundation (NSF) Cyber Trust grant CNS-0716335, and by Army Research Office (ARO) grant W911NF-07-1-0524.

Z. Kong and Edmund M. Yeh are with the Department of Electrical Engineering, Yale University (email: zhenning.kong@yale.edu, edmund.yeh@yale.edu).

E. Soljanin is with Bell Laboratories, Alcatel-Lucent, Murray Hill, NJ 07974, USA, (email: emina@lucent.com).

[1] We use the following notation. We say $f(n) = O(g(n))$ if there exists $n_0 > 0$ and a constant $M$ such that $|f(n)| \leq M|g(n)| \ \forall n \geq n_0$. We say $f(n) = o(g(n))$ if for any constant $\epsilon > 0$ there exists $n(\epsilon) > 0$ such that $|f(n)| \leq \epsilon|g(n)| \ \forall n \geq n(\epsilon)$. We say $f(n) = \Omega(g(n))$ if $g(n) = O(f(n))$, and $f(n) = \omega(g(n))$ if $g(n) = o(f(n))$. Finally, we say $f(n) = \Theta(g(n))$ if $f(n) = O(g(n))$ and $f(n) = \Omega(g(n))$.

all nodes follow a uniform, stationary and ergodic mobility model, the 2-hop relay scheme can achieve a constant, i.e., $\Theta(1)$, per S-D throughput, which does not vanish as the network size grows. Diggavi *et al.* show in [3] that a constant per S-D throughput can be achieved even with constrained mobility models.

The price paid for such an improvement on throughput is large delay. Characterizing the trade-off between the throughput and delay in mobile wireless networks has since attracted intense interest [4]–[11]. In [8], El Gamal *et al.* study the trade-off between per S-D throughput and delay in static and mobile wireless networks. They show that in static networks, the throughput-delay tradeoff satisfies $D(n) = \Theta(nT(n))$. Assuming a random walk mobility model, they show in [8] that the delay scaling of Grossglauser and Tse's 2-hop relay scheme is $\Theta(n \log n)$. The same throughput-delay scaling behavior is also obtained by Lin *et al.* in [10] for a Brownian motion mobility model. In [5], Neely and Modiano study the throughput-delay trade-off in mobile wireless networks under an i.i.d. mobility model, where they partition the unit square into $\Theta(n)$ cells and assume that after one time slot, each node moves from its current cell to another uniformly chosen cell. Neely and Modiano show that under this i.i.d. mobility model, the throughput-delay trade-off satisfies $D(n)/T(n) \geq \Theta(n)$. Specifically, they show that the delay scaling of Grossglauser and Tse's 2-hop relay scheme is $\Theta(n)$. A similar i.i.d. mobility model is also considered in [4]. Given that different mobility models and assumptions yield different throughput-delay trade-off behaviors, Sharma *et al.* propose a unified framework to capture the trade-off in a general setting in [9].

All previous work discussed thus far did not consider coding techniques. Recently, Ying *et al.* proposed joint coding-scheduling algorithms to improve the throughput-delay performance for mobile wireless networks using rate-less codes (e.g. Raptor codes) [12]. They study the random walk model where the unit torus is divided into $1/S^2$ sub-squares and at each time mobile nodes move from its current sub-square to one of its eight adjacent sub-squares. In this case, they show that the optimal throughput is $O(\sqrt{D(n)/n})$ when $S = o(1)$ and $D = \omega(|\log S|/S^2)$. They present a joint coding-scheduling algorithm that achieves the optimal throughput $O(\sqrt{D(n)/n})$ when $S = o(1)$ and $D$ is both $\omega(\max\{(\log^2 n)|\log S|/S^6, \sqrt[3]{n}\log n\})$ and $o(n/\log^2 n)$. Note however, that with $D = o(n/\log^2 n)$, the optimal throughput is $o(1/\log n)$, which is vanishing as $n \to \infty$. Hence the algorithms proposed in [12] are restricted to cases with small delay and small throughput, where a constant throughput ($\Theta(1)$) is not achievable. More importantly, when $S = 1/\sqrt{n}$, the model in [12] is identical to the one studied in [8]. However, the results of [12] do not apply to this case, since the set of feasible values of $D$ is empty.

In this paper, we study the throughput-delay trade-off in mobile wireless networks employing Reed-Solomon (RS) coding. In particular, we propose a 2-Hop Relay with Reed-Solomon Coding (2HRRSC) scheme, and show that while maintaining a constant $\Theta(1)$ throughput, 2HRRSC scheme can achieve a delay scaling of $\Theta(n)$ under the random walk mobility model studied in [8]. The improvement of the delay from $\Theta(n \log n)$ to $\Theta(n)$ under the random walk mobility model is significant. The intuitive idea behind this improvement is as follows. In the 2-hop relay scheme achieving $\Theta(1)$ throughput in [8], a particular relay node and a destination node need to meet and also be scheduled as a sender-receiver pair, which only occurs with a small probability. If they are not scheduled as a sender-receiver pair when they meet, they need to wait until they meet again and also be scheduled as a sender-receiver pair. This increases the packet delay. In contrast, in our scheme, the destination does not need to wait to meet a particular relay node, it can collect a packet from any relay nodes it encounters. This will be more clear as we present the detailed algorithm and analysis below.

The proposed 2HRRSC scheme not only provides a significant improvement in the network throughput-delay tradeoff, but also offers potential benefits in terms of robustness and security. In networks deployed in challenging environments, it is possible that a relay carrying a particular packet from the source never reaches the destination. In the 2-hop relay scheme without coding, this would mean that the packet would never be delivered to the destination. In the proposed 2HRRSC scheme, however, it is not required that all relays carrying coded versions reach the destination. As long as a sufficient fraction (to be specified below) of the versions are delivered, the destination can still decode all packets. Furthermore, in the 2HRRSC scheme, any relay carries at most a few coded versions of the packets. This means that the relays cannot typically decode the packets (which are intended for the destination) themselves. Hence, the proposed scheme is more secure than an uncoded scheme, where the relays have access to the uncoded packets they carry.

The rest of this paper is organized as follows. In Section II, we outline the network and mobility models that we use in this paper, and define some basic concepts. In Section III, we first briefly introduce Reed-Solomon (RS) codes, and then present a 2-Hop Relay with Reed-Solomon Coding (2HRRSC) scheme for mobile wireless networks. In Section VI we study the throughput-delay performance of the 2HRRSC scheme under the random walk mobility model. Finally, we conclude this paper in Section V.

## II. MODELS AND DEFINITIONS

Suppose that at time 0, $n$ nodes are uniformly distributed at random in the unit two-dimensional torus $\mathcal{B}$. Let $\mathbf{X}_u^{(t)}$ denote the location of node $u$ in $\mathcal{B}$ at time $t$. Each node serves as a source for one and only one destination, and each node serves as the destination for one and only one source. The network exhibits one possible source-destination pairing out of all possible pairings,[2] with all pairings being equally probable. Since each S-D pairings is equally probable, we assume throughout the paper that the network exhibits a particular pairing. Let S-D pair $i$ be the S-D pair with source node $i$ and its destination. Assume all sources have saturated transmission buffers, i.e., every source node always has a packet for its destination.

Divide the unit torus $\mathcal{B}$ into $\sqrt{n} \times \sqrt{n}$ equal-sized cells,[3] and label them as $\{(i,j) : i,j = 0,1,...,\sqrt{n}-1\}$. Assume that time is slotted such that each node moves from one cell to another cell after each time slot. In this paper, we consider a *random walk* mobility model where each node performs a simple random walk on the $\sqrt{n} \times \sqrt{n}$ grid. That is, if a node $u$ is in cell $(i,j)$ at time $t$, then at time $t+1$, the node $u$ is in any one of the four adjacent cells of $(i,j)$ with equal probability, and the location of node $u$, $\mathbf{X}_u^{(t+1)}$, is uniformly chosen at random in the new cell. By adjacent cells of $(i,j)$, we mean cells $\{(i+1,j),(i-1,j),(i,j+1),(i,j-1)\}$ with the addition and subtraction being modulo $\sqrt{n}$.

Under the network and mobility model described above, suppose a node $u$ transmits a packet at time $t$, then a node $v$ can receive this packet successfully if and only if for any other transmitting node $w$ in the network, $||\mathbf{X}_w^{(t)} - \mathbf{X}_v^{(t)}|| \geq (1+\Delta)||\mathbf{X}_u^{(t)} - \mathbf{X}_v^{(t)}||$, where $||\cdot||$ is the Euclidean distance and $\Delta$ is a positive number. This model is referred to as the *Protocol model* [1], and has been widely used in [4], [5], [8]–[10]. Another model for transmission is the *Physical model* [1]. Since these two models are essentially equivalent for the analysis of throughput and delay scaling performance in large-scale wireless networks [1], we focus on the Protocol model in this paper.

Transmissions in our mobile wireless network are coordinated and controlled by a scheduling scheme. More precisely, a scheduling scheme $\pi$ is a sequence of polices $\{\pi_k\}$ which determines which nodes in the network transmit, and which packet is transmitted at each node, at each time slot $k = 1, 2, ....$ For a given scheduling scheme, the throughput and delay are defined as follows:

*Definition 1:* (Throughput) For a given scheduling scheme $\pi$, let $H_\pi(i,t)$ be the total number of packets

---
[2] The total number of pairings is $\sum_{k=0}^{n}(-1)^k \binom{n}{k}(n-k)!$.
[3] For simplicity, we ignore integer constraints throughout the paper.

transmitted for S-D pair $i$ up to time $t$. The long-term throughput for S-D pair $i$ is

$$\liminf_{t \to \infty} \frac{1}{t} H_\pi(i, t).$$

The average throughput over all S-D pairs is

$$T'_\pi(n) = \frac{1}{n} \sum_{i=1}^{n} \liminf_{t \to \infty} \frac{1}{t} H_\pi(i, t).$$

The throughput of $\pi$, $T_\pi(n)$, is defined as the expectation over all network realizations of the average throughput over all S-D pairs, i.e.,

$$T_\pi(n) \triangleq E[T'_\pi(n)]. \tag{1}$$

Note that a network realization includes a realization of initial node positions $\{\mathbf{X}_1^{(0)}, ..., \mathbf{X}_n^{(0)}\}$ and node movement trajectories $\{\mathbf{X}_1^{(t)}, ..., \mathbf{X}_n^{(t)}\}$ for all $t \geq 0$.

*Definition 2:* (Delay) For a given scheduling scheme $\pi$, let $D_\pi(i, k)$ be the delay of packet $k$ (from the time the packet starts transmission at the source until the time the packet is decoded successfully at the destination) for S-D pair $i$. The delay for S-D pair $i$ is

$$\limsup_{h \to \infty} \frac{1}{h} \sum_{k=1}^{h} D_\pi(i, k).$$

The average delay over all S-D pairs is

$$D'_\pi(n) = \frac{1}{n} \sum_{i=1}^{n} \limsup_{h \to \infty} \frac{1}{h} \sum_{k=1}^{h} D_\pi(i, k).$$

The delay of $\pi$, $D_\pi(n)$, is defined as the expectation over all network realizations of the average delay over all S-D pairs, i.e.,

$$D_\pi(n) \triangleq E[D'_\pi(n)]. \tag{2}$$

The throughput $T'_\pi(n)$ and delay $D'_\pi(n)$ are both random variables, since they depend on the initial node locations and random node movements. The throughput $T_\pi(n)$ and delay $D_\pi(n)$ are ensemble averages. To study the asymptotical behavior of $T_\pi(n)$ and $D_\pi(n)$, we will let the number of nodes $n$ go to infinity. We say an event holds *with high probability* (w.h.p.) if the event occurs with probability 1 as $n$ goes to infinity.

### III. 2-HOP RELAY WITH REED-SOLOMON CODING SCHEME

In this section, we first briefly describe the Reed-Solomon (RS) coding scheme, and then propose a 2-hop relaying scheme with RS codes that achieves $\Theta(1)$ throughput and $\Theta(n)$ delay under the random walk mobility model.

*A. Reed-Solomon Codes*

An $(n, m)$ linear code $\mathcal{C}$ over a finite field $\mathbb{F}_q$ is an $m$-dimensional subspace of the vector space $\mathbb{F}_q^n$ of all $n$-tuples over $\mathbb{F}_q$, consisting of $q^m$ codewords. Here $q = p^s$ is the size of the finite field $\mathbb{F}_q$, where $p$ is a prime number and $s$ is an integer. An $(n, m)$ linear code with minimum Hamming distance $d$ is called an $(n, m, d)$ linear code.

A fundamental relationship for the parameters $(n, m, d)$ over any field is the Singleton bound: $d \leq n - m + 1$. Any code that satisfies the Singleton bound with equality is called a Maximum Distance Separable (MDS) code. A useful class of MDS codes are the Reed-Solomon (RS) codes. It is known that for any code parameters $n, m, d = n - m + 1, (1 \leq m \leq n)$ and finite field $\mathbb{F}_q$, there exists a linear $(n, m, d)$ (extended) RS code over $\mathbb{F}_q$, as long as $n \leq q + 1$ [13]. A natural way to define RS codes is to take $n = q$. Let the $m$ information symbols be denoted by $(y_0, y_1, \ldots, y_{m-1})$ where $y_i \in \mathbb{F}_q, 0 \leq i \leq m - 1$. Let $f(x) = y_0 + y_1 x + \cdots + y_{m-1} x^{m-1}$ be the corresponding polynomial in the indeterminate $x$. Let $\beta_1, \beta_2, ..., \beta_q$ be the $q$ different elements of $\mathbb{F}_q$ arranged in some order. The information polynomial $f(x)$ is mapped into the $q$-tuple $(f(\beta_1), f(\beta_2), ..., f(\beta_q))$ over $\mathbb{F}_q$, where $f(\beta_i)$ is equal to the polynomial $f(x)$ evaluated at element $\beta_i \in \mathbb{F}_q$:

$$f(\beta_i) = \sum_{j=1}^{m-1} y_j \beta_i^j, \quad 1 \leq i \leq q. \tag{3}$$

The $q^m$ $q$-tuples generated by the mapping $f(x) \to \{f(\beta_i), \beta_i \in \mathbb{F}_q\}$ as the polynomial $f(x)$ ranges over all $q^m$ polynomials over $\mathbb{F}_q$ of degree strictly less than $m$ form a linear $(n = q, m, d = n - m + 1)$ MDS code over $\mathbb{F}_q$. This code is called a Reed-Solomon (RS) code.

An essential property of RS codes (or any MDS code) is that every subset of $m = n - d + 1$ coordinates is an *information set* of $\mathcal{C}$, in the sense that the codewords run through all $q^m$ possible $m$-tuples in any subset of $m$ coordinates. Thus, the $m$ information symbols can be recovered from *any* $m = n - d + 1$ received coded symbols. This is crucial for the improvement of the throughput-delay trade-off when we apply this coding technique to mobile wireless networks. A series of efficient RS decoding algorithms by Peterson, Berlekamp-Massey, Euclid, Welch-Berlekamp and Sudan have been developed over the years. For details, please refer to [13].

*B. Coding and Relaying Algorithms*

As we explained in the network model, each node serves as a source for one and only one destination, and each destination has one and only one source. Consequently, there are $n$ sessions (S-D pairs) in the network. In addition, each node serves as a relay for all the other $n-1$ S-D pairs. Due to these different roles, a node has different encoding/decoding and packet relaying operations.

We treat each packet as a symbol over a finite field $\mathbb{F}_q$ with $q = p^s$ for a prime number $p$ and an integer $s$. We require that $2^l \leq q$, where $l$ is the length of each packet. Furthermore, it is required that $n \leq q+1$. This implies that as $n$ gets large, the size of the finite field has to grow at least with $n$. We assume these relationships hold throughout this paper.

(i) *Source Encoding*: Each node $u$ first groups every $m$ source packets into one generation for its destination. In other words, for S-D pair $u$, each generation $g = 1, 2, ...$ consists of $m$ consecutive packets as an $m$-tuple $\mathbf{y}^{u,g} = (y_1^{u,g}, y_2^{u,g}, ..., y_m^{u,g}) \in (\mathbb{F}_q)^m$. Node $u$ then applies the RS encoding algorithm to generate a "codeword" as an $n$-tuple $(z_1^{u,g}, z_2^{u,g}, ..., z_n^{u,g}) \in (\mathbb{F}_q)^n$ (cf. (3)). We call $z_k^{u,g}$ the $k$th *version* of generation $g$, $k = 1, 2, ..., n$. The versions for one generation are stored in the order of version index $k$ in a buffer (assumed to have infinite capacity) for node $u$ designated for its destination.

(ii) *Relay Storing*: Each node has $n-1$ buffers for the other $n-1$ S-D pairs, and all buffers have infinite capacities. When a node $w$ serves as a relay for S-D pair $u$ and receives a version $z_k^{u,g}$ for S-D pair $u$, it puts this version into the buffer queue for S-D pair $u$.

(iii) *Destination Decoding*: As soon as the destination node of node $u$ receives any $m$ distinct versions for the same generation $g$, the destination node can recover the original source packets $y_1^{u,g}, y_2^{u,g}, ..., y_m^{u,g}$ by employing the RS decoding algorithm.

In this paper, we do not consider the delay incurred by the encoding and decoding processes.

*C. 2-Hop Relay with Reed-Solomon Coding (2HRRSC) Scheme*

In our 2-hop relay scheme, we allow nodes to transmit only to nearby nodes. Precisely, a node $u$ in a cell $(i, j)$ may transmit only to the other nodes in the same cell $(i, j)$.

*Definition 3:* (2-Hop Relay with RS Coding (2HRRSC) Scheme) We assume that there is a hand-shaking protocol that allows each node to identify those nodes in its cell, the generations that those nodes have already successfully decoded, and the versions which have been received for un-decoded generations.[4]

When a node $u$ is scheduled as a sender at time $t$ (the scheduling scheme will be discussed after Lemma 1), it chooses one of the following two actions:

(i) Serves as source with probability $p_s$: If there is at least one node other than $u$ in $u$'s cell, node $u$ chooses one such node $v$, uniformly at random, and transmits the HOL (head-of-line) version of the current generation for $u$'s destination to node $v$. If node $v$ is not $u$'s destination, $v$ performs the relay storing algorithm. If there are no other nodes in $u$'s cell, nothing happens. If all $n$ versions of the current generation $g$ have been sent out, node $u$ moves to the next generation $g+1$ and applies the source encoding algorithm.

(ii) Serves as relay with probability $1 - p_s$: If there are nodes other than $u$ in $u$'s cell, node $u$ chooses one such node $v$, uniformly at random. Suppose that node $v$ has successfully decoded generations $1, 2, ..., g$, then node $u$ transmits the head-of-line (HOL) innovative version for the earliest (having the smallest index) undecoded generation (in this case $g+1$) in $u$'s buffer to node $v$. Here, an innovative version is a version that node $v$ does not yet have. Node $u$ also discards all versions for decoded generations $1, 2, ..., g$, if it has any. If node $u$ does not have an innovative version for an undecoded generation for $v$, or if there are no other nodes in $u$'s cell, nothing happens.

In the following section, we will show that this 2HRRSC scheme can achieve $\Theta(1)$ (average) S-D throughput and $\Theta(n)$ (average) packet delay. The improvement of the delay from $\Theta(n \log n)$ to $\Theta(n)$ relies on the use of RS coding. To see this, note that in the 2-hop relay scheme achieving $\Theta(1)$ throughput in [8], a particular relay node $w$ and a destination node $v$ need to meet and be scheduled as a sender-receiver pair, which only occurs with a small probability. If $w$ and $v$ are not scheduled as a sender-receiver pair, they need to wait until they meet again and be scheduled as a sender-receiver pair. This increases the delay for the packet carried by the node $w$. In contrast, in our scheme with RS coding, the destination does not need to wait to meet a particular relay node. It can collect a version from any relay node it encounters, and can perform decoding as soon as it collects any $m$ out of the $n$ distinct versions.

---

[4]Protocols similar to the hand-shaking protocol discussed here are also implicitly assumed for all the networks studied in [4]–[11]. For simplicity, we do not consider the overhead associated with this protocol.

# IV. Throughput-Delay Performance of 2HRRSC

In this section, we study the throughput-delay performance of the proposed 2HRRSC scheme. We show that the 2HRRSC scheme can achieve $\Theta(1)$ throughput and $\Theta(n)$ delay under the random walk mobility model.

## A. Scheduling Scheme

We say that a node $u$ has achieved a *successful transmission* if $u$ transmits (either as a source or as a relay) a packet and the receiver successfully receives the packet under the Protocol model.

*Lemma 1:* Let $\theta(n)$ be the fraction of nodes that can have successful transmission simultaneously at any given time. Then, there exists a scheduling scheme such that

$$E[\theta(n)] = \theta' \left[ 1 - \left(1 - \frac{1}{n}\right)^{n-1} - \left(1 - \frac{1}{n}\right)^n \right], \quad (4)$$

where $\theta' = (\lceil \sqrt{2}(1+\Delta) \rceil + 1)^{-2}$ and $\Delta$ is the parameter defined in the Protocol model.

*Proof:* We first show that among the $n$ cells, there exist $\theta' n$ cells, where $\theta'$ is a positive constant independent of $n$, such that simultaneous transmissions in these cells can take place without interfering with each other. This is similar to the result proved in [8].

Consider a node $u$ in a cell $(i,j)$. The largest possible distance between $u$ and any other node in its cell is $\sqrt{2}d$, where $d = 1/\sqrt{n}$ is the edge length of the cell. Under the Protocol Model, if all other transmitting nodes are away from node $u$'s cell by a distance greater than or equal to $\lceil (1+\Delta)\sqrt{2} \rceil d$, then node $u$'s transmission can be received successfully. Thus, as shown in Fig. 1, the fraction of cells that can be chosen to have simultaneous transmissions is at least

$$\theta' \triangleq \frac{d^2}{(\lceil (1+\Delta)\sqrt{2} \rceil d + d)^2} = \frac{1}{(\lceil \sqrt{2}(1+\Delta) \rceil + 1)^2}. \quad (5)$$

We refer to the cells which can have simultaneous transmissions as active cells.

Given that a cell is active, a successful transmission can take place in the cell if there are at least two nodes in the cell. Let $p_1(n)$ be the probability that there are at least two nodes in a cell. Since the initial positions $\mathbf{X}_1^{(0)}, \mathbf{X}_2^{(0)}, ..., \mathbf{X}_n^{(0)}$ are uniformly distributed at random in the unit torus, the positions $\mathbf{X}_1^{(t)}, \mathbf{X}_2^{(t)}, ..., \mathbf{X}_n^{(t)}$ are i.i.d. and uniform at any given time $t \geq 0$ under the random walk mobility model. Thus, the number of nodes in each cell has a binomial distribution with mean value $1$. Therefore, we have

$$p_1(n) = \left[ 1 - \left(1 - \frac{1}{n}\right)^n - \left(1 - \frac{1}{n}\right)^{n-1} \right]. \quad (6)$$

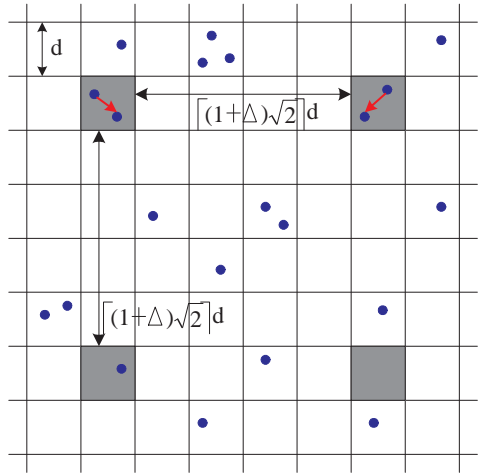

Fig. 1. Scheduling of active cells

Thus, the mean number of nodes that can be scheduled to have successful transmissions simultaneously is $p_1(n)\theta'n$. The mean fraction of nodes that can be scheduled to have successful transmissions simultaneously is $E[\theta(n)] = \theta' p_1(n)$, and we obtain (4). As $n \to \infty$, we have

$$\theta_0 \triangleq \lim_{n \to \infty} E[\theta(n)] = \theta'(1 - 2e^{-1}). \tag{7}$$

$\square$

Lemma 1 asserts that, at any given time, a positive expected fraction $E[\theta(n)]$ of the $n$ nodes can have successful transmissions simultaneously. To achieve this, we need to first schedule active cells and then schedule transmitting nodes within the active cells. For the first part, according to the proof of Lemma 1, any given cell can be scheduled to be active every $\frac{1}{\theta'}$ time slots. This is the approach used in [8]. In fact, since the nodes are mobile, we can fix a set of cells with a regular pattern (e.g., the shaded cells in Fig. 1) which are scheduled to be always active. Then, consider the following scheduling scheme for transmitting nodes: when there are at least two nodes in an active cell, one node, chosen uniformly at random, is scheduled as a sender. The sender then chooses one node among the remaining nodes in the active cell uniformly at random as the receiver.

We can show that the probability that any given node has a successful transmission is also $E[\theta(n)]$ as follows. The probability that a given node $u$ is in an active cell is $\theta'$. When there are $k \geq 1$ nodes other than $u$ in the same active cell as $u$, the probability that $u$ is chosen as a sender is $\frac{1}{k+1}$. Hence, the probability that node $u$ is scheduled as a sender is

$$\theta' \sum_{k=1}^{n-1} \frac{1}{k+1} \binom{n-1}{k} \left(\frac{1}{n}\right)^k \left(1-\frac{1}{n}\right)^{n-1-k} = \theta' p_1(n) = E[\theta(n)]. \tag{8}$$

Let $Y$ be the time between two successful transmissions for any given node. Because the underlying mobility process and the scheduling scheme are both Markov, we have

$$E[Y] = \frac{1}{E[\theta(n)]}. \tag{9}$$

*B. Packet Delay*

We say that two nodes *meet* if they are in the same cell. Under the random walk mobility model, since each node independently follows a simple random walk, the joint position of the two nodes can be viewed as a difference random walk with respect to the position of one node. The inter-meeting times are just the inter-entering times to $(0,0)$ for the difference random walk on a $\sqrt{n} \times \sqrt{n}$ grid. Then, for any two nodes $u$ and $v$, the inter-meeting times are i.i.d. The following lemma provided in [8] gives the mean and variance of the inter-meeting time $\tau$.

*Lemma 2:* [8] Under the random walk mobility model, the expected value and the variance of the inter-meeting time $\tau$ are given by

$$E[\tau] = n, \quad \text{Var}(\tau) = \Theta(n^2 \log n). \tag{10}$$

Note that $Y$ is the time between two successful transmissions for a given node, where the role (i.e., source or relay) of the transmitting node is not specified. Now let $Y'$ be the time between two successful transmissions for a node acting as a source. Then $Y' = Y_1 + Y_2 + \cdots + Y_{H'}$, where $H'$ is a geometric random variable with parameter $p_s$. Then

$$E[Y'] = E[H']E[Y] = \frac{1}{p_s E[\theta(n)]} = \frac{1}{p_s \theta' p_1(n)}. \tag{11}$$

In the 2HRRSC scheme, every source transmits one generation $n$ times ($n$ distinct versions). Nevertheless, some nodes may receive multiple distinct versions and some others may not receive any version. The following lemma asserts that at a certain time instant after the source node sends out the first version of a given generation, there exists a constant fraction of nodes in the network carrying versions of the given generation.

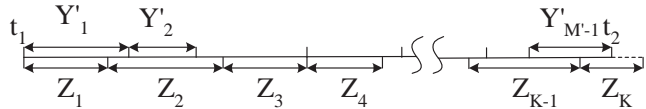

Fig. 2. Time instants of transmission by node $u$, and inter-meeting times of nodes $u$ and $w_1$.

*Lemma 3:* In the 2HRRSC scheme, let $t_1$ be the time instant when source node $u$ transmits the first version of generation $g$. Let

$$t_s(n) = \frac{n}{E[\theta(n)]p_s} = \frac{n}{p_s\theta'p_1(n)}, \quad (12)$$

and $M$ be the number of nodes that receive at least one version of generation $g$ from $u$ during $[t_1, t_2]$, where $t_2 = t_1 + t_s(n)$ and $[t_1, t_2]$ refers to the set of time slots $\{t_1, t_1+1, ..., t_2 = t_1 + t_s(n)\}$. Let $\phi(n) = \frac{M}{n}$. Then, for any $0 < \epsilon < 1$, we have

$$\Pr\left(\phi(n) \leq \frac{1-\epsilon}{\frac{1}{p_s\theta'p_1(n)}+1}\right) = \Theta\left(\frac{\log n}{n}\right). \quad (13)$$

*Proof:* Let $M'$ be the number of versions that the source $u$ sends out during $[t_1, t_2]$. Then there are $M'$ receivers receiving the versions, although the corresponding nodes are not necessarily distinct. Since in any time slot, the probability that node $u$ has a successful transmission as a source is $p_s E[\theta(n)] = p_s\theta'p_1(n)$, we have

$$E[M'] = t_s(n)p_s\theta'p_1(n) = n. \quad (14)$$

Suppose node $u$ transmits a version to node $w_1$ at $t_1$. Let $Z_i$ be the time between the $i$th and the $(i+1)$th meetings between $u$ and $w_1$, as illustrated in Fig. 2. Note that when $w_1$ and $u$ meet again, node $u$ may not be scheduled as a sender.

Now define $K \triangleq \min\{j \in \mathbb{N} : \sum_{i=1}^{j} Z_i > t_s(n)\}$. In other words, $K$ is the number of times node $u$ meets $w_1$ during $[t_1, t_2]$. By definition, we have

$$E\left[\sum_{i=1}^{K-1} Z_i\right] \leq t_s(n), \quad \text{and} \quad E\left[\sum_{i=1}^{K} Z_i\right] > t_s(n).$$

Since the $Z_i$'s are i.i.d. and $K$ is a stopping time for $\{Z_i\}$, by Wald's equality [14], we have

$$E\left[\sum_{i=1}^{K} Z_i\right] = E[Z_i]E[K],$$

and

$$E\left[\sum_{i=1}^{K-1} Z_i\right] = (E[K]-1)E[Z_i].$$

Since $E[Z_i] = E[\tau] = n$, we have

$$\frac{t_s(n)}{n} < E[K] \leq \frac{t_s(n)}{n} + 1, \tag{15}$$

or

$$\frac{1}{p_s \theta' p_1(n)} < E[K] \leq \frac{1}{p_s \theta' p_1(n)} + 1. \tag{16}$$

Moreover, by Theorem 10 in Appendix A we have[5]

$$\text{Var}(K) = \frac{t_s(n)\text{Var}(Z_i)}{E[Z_i]^3} + o(1) = \frac{\frac{n}{p_s \theta' p_1(n)} \Theta(n^2 \log n)}{n^3} + o(1) = \Theta(\log n). \tag{17}$$

Let $M$ be the number of distinct nodes $\{w_1, w_2, ..., w_M\}$ among those $M'$ receivers that receive a version from $u$ during $[t_1, t_2]$. Let $R_j$ be the number of times that node $u$ meets node $w_j$ during $[t_1, t_2]$, where $j = 1, 2, ..., M$. By coupling methods, it is easy to show that $E[R_j] \leq E[K]$.

Since $M' = \sum_{j=1}^{M} R_j$, we have,

$$E[M'] = E\left[E\left[\sum_{j=1}^{M} R_j | M\right]\right] \leq E[M]E[K], \tag{18}$$

which yields

$$E[M] \geq \frac{E[M']}{E[K]} \geq \frac{n}{\frac{1}{p_s \theta' p_1(n)} + 1}. \tag{19}$$

Since $M \leq n$, we have $E[M] = \Theta(n)$.

Now consider i.i.d. random variables $K_j, j = 1, ..., M$ having the same distribution as $K$. Note that

$$\text{Var}\left(\sum_{j=1}^{M} K_j\right) \leq \text{Var}\left(\sum_{j=1}^{n} K_j\right) = n\text{Var}(K) = \Theta(n \log n). \tag{20}$$

On the other hand, by the law of total variance, we have

$$\begin{aligned}
\text{Var}\left(\sum_{j=1}^{M} K_j\right) &= E\left[\text{Var}\left(\sum_{j=1}^{M} K_j | M\right)\right] + \text{Var}\left(E\left[\sum_{j=1}^{M} K_j | M\right]\right) \\
&= E\left[\text{Var}\left(\sum_{j=1}^{M} K_j | M\right)\right] + \text{Var}(ME[K]) \\
&= E\left[\text{Var}\left(\sum_{j=1}^{M} K_j | M\right)\right] + \text{Var}(M)E[K]^2 \\
&= E\left[\text{Var}\left(\sum_{j=1}^{M} K_j | M\right)\right] + \text{Var}(M)\Theta(1). \tag{21}
\end{aligned}$$

---

[5]Here, in the context of Theorem 10, $b = t_s(n)$, $\sigma^2 = \text{Var}(Z_i) = \text{Var}(\tau) = \Theta(n^2 \log n)$ and $\mu = E[Z_i] = E[\tau] = n$.

By (20) and (21), we have

$$\text{Var}(M) = O(n \log n). \quad (22)$$

Thus

$$E[\phi(n)] = \frac{E[M]}{n} \geq \frac{1}{\frac{1}{p_s \theta' p_1(n)} + 1}, \quad (23)$$

and

$$\text{Var}(\phi(n)) = \frac{1}{n^2} \text{Var}(M) = O\left(\frac{\log n}{n}\right). \quad (24)$$

By Chebyshev inequality, for any $0 < \epsilon < 1$,

$$\Pr\left(\phi(n) \leq \frac{1-\epsilon}{\frac{1}{p_s \theta' p_1(n)} + 1}\right) \leq \Pr\left(\phi(n) \leq (1-\epsilon)E[\phi(n)]\right)$$
$$\leq \frac{\text{Var}(\phi(n))}{\epsilon^2 E[\phi(n)]^2}$$
$$\leq \frac{\text{Var}(\phi(n))}{\epsilon^2 \left(\frac{1}{p_s \theta' p_1(n)} + 1\right)^{-2}}$$
$$= O\left(\frac{\log n}{n}\right).$$

This implies that $\phi(n) \geq (1-\epsilon)\frac{\theta_0 p_s}{1+\theta_0 p_s}$ for any $0 < \epsilon < 1$ w.h.p. □

Lemma 3 asserts that $t_s(n) = \frac{n}{p_s \theta' p_1(n)}$ time slots after the source node transmits the first version of a given generation, with high probability, there exist $\phi(n)n = \Theta(n)$ nodes having at least one version of the given generation in the network. How these nodes and the destination node are distributed in the network affects how fast the destination node can collect $m$ distinct versions to decode a given generation. Due to the nature of the random walk mobility model, we know that given any initial distribution, after some time, the distribution of each node converges to the uniform distribution. This effect can be captured by the *mixing time*. More precisely, we define the $\epsilon$-mixing time as follows.

*Definition 4:* Let $P = [P_{ij}]$ be the transition matrix of a Markov chain with uniform stationary distribution. Let $P^t$ be the $t$th power of $P$. Define

$$\Delta_i(t) \triangleq \sum_{j=1}^{n} \left| P_{ij}^t - \frac{1}{n} \right|. \quad (25)$$

Then, the $\epsilon$-mixing time is defined as

$$T_{mix}(P, \epsilon) \triangleq \sup_i \inf \left\{ t : \Delta_i(t') \leq \epsilon \text{ for all } t' > t \right\}. \quad (26)$$

*Lemma 4:* Let $P$ be the transition matrix of a random walk on a two-dimensional torus. Let $T_{mix}(P, \epsilon)$ be defined as in (26), then

$$\frac{2 \log n - 1}{n - 1} \leq T_{mix}(P, 1/n^2) \leq \frac{3 \log n}{1 - \frac{1}{n}}. \quad (27)$$

*Proof:* Due to the symmetric nature of the random walk mobility model, the transition matrix $P$ for each node is a doubly stochastic matrix. Let $1 = \lambda_1(P) \geq \lambda_2(P) \geq \cdots \geq \lambda_n(P) \geq -1$ be the eigenvalues of $P$. Let $\lambda_{max}(P) = \max\{\lambda_2(P), -\lambda_n(P)\}$. The following result gives well-known bounds on $T_{mix}(P, \epsilon)$ [15].

$$\frac{\lambda_{max}(P) \log(2\epsilon)^{-1}}{1 - \lambda_{max}(P)} \leq T_{mix}(P, \epsilon) \leq \frac{\log n + \log \epsilon^{-1}}{1 - \lambda_{max}(P)}. \quad (28)$$

For a random walk on two-dimensional torus, $\lambda_{max}(P) = \frac{1}{n}$ [16]. Substituting $\lambda_{max}(P) = \frac{1}{n}$ and $\epsilon = \frac{1}{n^2}$, we obtain (27). □

Suppose the distribution of node $u$ among the $n$ cells at time $t_0$ is $\mathbf{f}^{(t_0)} = (f_1^{(t_0)}, f_2^{(t_0)}, ..., f_n^{(t_0)})$, then after $t$ time slots, the distribution is $\mathbf{f}^{(t+t_0)} = \mathbf{f}^{(t_0)} P^t$, where $P$ is the transition matrix. The probability of state $i$ is $f_i^{(t+t_0)} = \sum_{k=1}^n f_k^{(t_0)} P_{ki}^t$. Then we have

$$\begin{aligned}
\left| f_i^{(t+t_0)} - \frac{1}{n} \right| &= \sum_{k=1}^n f_k^{(t_0)} \left| P_{ki}^t - \frac{1}{n} \right| \\
&\leq \sum_{k=1}^n \left| P_{ki}^t - \frac{1}{n} \right| \\
&= \Delta_i(t).
\end{aligned}$$

Hence, by the definition of $T_{mix}(P, 1/n^2)$ and Lemma 4, after $c_0 \log n$ time slots for some constant $c_0 \geq 3/(1 - \frac{1}{n})$, the probability that any given node $u$ is in any given cell $(i, j)$ is bounded between $\frac{1}{n} - \frac{1}{n^2}$ and $\frac{1}{n} + \frac{1}{n^2}$. That is, the distribution is approximately uniform. This observation leads to the following lemma, for which the proof is given in Appendix B.

*Lemma 5:* Suppose that $t_s(n) = \frac{n}{p_s \theta' p_1(n)}$ time slots after the source node transmits the first version of generation $g$, there exist $M$ distinct nodes having versions of generation $g$. Let $\eta(n)$ be the probability that a scheduled sender has a version of generation $g$ after another $c_0 \log n$ time slots where $c_0 \geq 3/(1 - \frac{1}{n})$. Then $\eta(n) = \phi(n) = \frac{M}{n}$ w.h.p.

In the 2HRRSC scheme, when a relay node $w$ is scheduled to transmit, it chooses one of the nodes in its cell, uniformly at random, as a destination node, and transmits a version designated for that node.

Similar to Lemma 1, the following lemma quantifies the probability that (in any time slot) node $v$ is chosen as the receiver by a scheduled sender which performs as a relay for $v$.

*Lemma 6:* Let $p_2(n)$ be the probability that (in any time slot) node $v$ is chosen as the receiver by a scheduled sender which performs as a relay for $v$. Then

$$p_2(n) = \theta'(1-p_s)\frac{n-2}{n-1}\left[1-\left(1-\frac{1}{n}\right)^n-\left(1-\frac{1}{n}\right)^{n-1}\right]. \qquad (29)$$

*Proof:* To be scheduled as a receiver, node $v$ needs to be in an active cell, which occurs with probability $\theta'$. Assume there are $k \geq 1$ nodes other than $v$ in the active cell, then the probability that $v$ is scheduled as a receiver is $\left(1-\frac{1}{k+1}\right)\frac{1}{k}$. Furthermore, the probability that the scheduled sender is not the source and performs as a relay for node $v$ is $\frac{n-2}{n-1}(1-p_s)$. Therefore, we have

$$\begin{aligned}
p_2(n) &= \theta'\sum_{k=1}^{n-1}(1-p_s)\frac{n-2}{n-1}\left(1-\frac{1}{k+1}\right)\frac{1}{k}\binom{n-1}{k}\left(\frac{1}{n}\right)^k\left(1-\frac{1}{n}\right)^{n-1-k} \\
&= \theta'\sum_{k=1}^{n-1}(1-p_s)\frac{n-2}{n-1}\frac{1}{k+1}\binom{n-1}{k}\left(\frac{1}{n}\right)^k\left(1-\frac{1}{n}\right)^{n-1-k} \\
&= \theta'(1-p_s)\frac{n-2}{n-1}\sum_{k=1}^{n-1}\frac{1}{k+1}\frac{(n-1)!}{(n-1-k)!k!}\left(\frac{1}{n}\right)^k\left(1-\frac{1}{n}\right)^{n-1-k} \\
&= \theta'(1-p_s)\frac{n-2}{n-1}\sum_{k=1}^{n-1}\frac{n!}{[n-(k+1)]!(k+1)!}\left(\frac{1}{n}\right)^{k+1}\left(1-\frac{1}{n}\right)^{n-(k+1)} \\
&= \theta'(1-p_s)\frac{n-2}{n-1}\sum_{k=1}^{n-1}\binom{n}{k+1}\left(\frac{1}{n}\right)^{k+1}\left(1-\frac{1}{n}\right)^{n-(k+1)} \\
&= \theta'(1-p_s)\frac{n-2}{n-1}\sum_{k=2}^{n}\binom{n}{k}\left(\frac{1}{n}\right)^{k}\left(1-\frac{1}{n}\right)^{n-k} \\
&= \theta'(1-p_s)\frac{n-2}{n-1}\left[1-\left(1-\frac{1}{n}\right)^n-\left(1-\frac{1}{n}\right)^{n-1}\right] \\
&= \theta'(1-p_s)\frac{n-2}{n-1}p_1(n). \qquad (30)
\end{aligned}$$

As $n \to \infty$,

$$p_2 \triangleq \lim_{n\to\infty} p_2(n) = \theta'(1-p_s)(1-2e^{-1}). \qquad (31)$$

□

Based on the above results, the following lemma asserts that when $m$—the number of packets in one generation—is properly chosen, the destination node can collect $m$ distinct versions for that generation

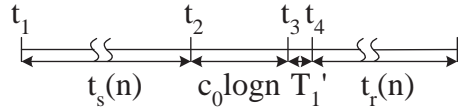

Fig. 3. Key time instants and duration for transmission and collection analysis.

within $\Theta(n)$ time slots w.h.p.

*Lemma 7:* Consider a particular S-D pair with source node $u$ and destination node $v$. For any positive constant $c_1$, there exists a positive constant $\delta$ satisfying

$$\delta < \frac{p_s \theta_0}{1 + \theta_0 p_s} \cdot \frac{p_2 c_1}{c_1 + 1}, \tag{32}$$

such that by choosing the number of packets in a generation to be $m = \delta n$, node $v$ can collect $m$ distinct versions for the given generation within $c_1 n$ time slots w.h.p.

*Proof:* At $t_s(n) = \frac{n}{p_s \theta' p_1(n)}$ time slots after the source node $u$ sends out the first version for a given generation, there exist $M = \phi(n) n$ nodes that have at least one version of the given generation. In order to decode that generation, the destination node $v$ needs to collect $m$ versions. Let the time required for accomplishing this be $T(n)$.

As defined in Lemma 5, $t_1$ is the time instant when $u$ sends out the first version of the given generation, and $t_2 = t_1 + t_s(n)$. Consider a virtual system in which the destination node $v$ knows $t_1$ and $t_2$. In this virtual system, instead of collecting versions for the given generation as soon as possible, $v$ waits another $c_0 \log n$ time slots after $t_2$, where $c_0 \geq 3/(1 - \frac{1}{n})$, and then begins to collect versions. Let $\tilde{T}(n)$ be the time required for node $j$ to collect $m$ versions of the given generation in the virtual system. Clearly, $\tilde{T}(n)$ stochastically upper bounds $T(n)$, i.e., $\Pr(\tilde{T}(n) > x) \geq \Pr(T(n) > x)$, for all $x > 0$. In the following, we consider this virtual system.

Suppose $w_1$ is the first relay node that transmits successfully to $v$ after $t_3 = t_2 + c_0 \log n$, and the time instant when $w_1$ transmits to $v$ is $t_4 = t_3 + T'_1$. Note that $T'_1$ is stochastically dominated by $Y''$, which is the time between successful transmissions where node $v$ is chosen as a receiver by scheduled senders which perform as relays for $v$. By the Markov inequality and (29), for any $c > 0$,

$$\Pr(Y'' < c \log n) > 1 - \frac{E[Y'']}{c \log n} = 1 - \Theta\left(\frac{1}{\log n}\right),$$

which indicates that $Y'' = o(\log n)$ w.h.p. Therefore, $T'_1 = o(\log n)$ w.h.p.

Let $t_r(n) = c_1 n - (c_0 + 1) \log n$, where $c_1 > \frac{c_0+1}{e \ln 2}$. Since $\max_{x>0} \frac{\log x}{x} = \frac{1}{e \ln 2}$, $t_r(n) > 0$ for any $n > 0$. The time instants $t_1, t_2, t_3, t_4$, and duration $t_s(n), T'_1, t_r(n)$ are illustrated in Fig. 3. Let $N'$ be the number of times that node $v$ is chosen as the receiver by a relay node during $[t_4, t_4 + t_r(n)]$. Since the probability that (in any time slot) node $v$ is chosen as the receiver by a scheduled sender which performs as a relay for $v$ is $p_2(n)$, we have

$$E[N'] = t_r(n) p_2(n), \tag{33}$$

which implies $E[N'] = \Theta(n)$.

Let $K'$ be the number of times that $w_1$ transmits successfully to $v$ during $[t_4, t_4 + t_r(n)]$. Then, as in the proof for Lemma 3, $E\left[\sum_{i=1}^{K'-1} Z_i\right] \leq t_r(n)$ and $E\left[\sum_{i=1}^{K'} Z_i\right] > t_r(n)$, where $Z_i, i = 1, 2, ..., K'$, is the time between the $i$th and the $(i+1)$th times when $w_1$ transmits successfully to $v$. By Wald's equality and the fact that the $Z_i$'s are i.i.d. with mean $E[\tau] = n$, we have

$$\frac{t_r(n)}{n} < E[K'] \leq \frac{t_r(n)}{n} + 1. \tag{34}$$

Hence $E[K'] = \Theta(1)$.

Moreover, by Theorem 10 in Appendix A we have[6]

$$\text{Var}(K') = \frac{t_r(n) \text{Var}(Z_i)}{E[Z_i]^3} + o(1) = \frac{[c_1 n - (c_0 + 1) \log n] \Theta(n^2 \log n)}{n^3} + o(1) = \Theta(\log n). \tag{35}$$

Let $N$ be the number of distinct nodes $\{w_1, w_2, ..., w_N\}$ among those $N'$ nodes that choose $v$ as the receiver during $[t_4, t_4 + t_r(n)]$. Let $R'_i$ be the number of times that $v$ is chosen as the target receiver by $w_i$ during $[t_4, t_4 + t_r(n)]$, where $i = 1, 2, ..., N$. Then $E[R'_i] \leq E[K']$. Since $N' = \sum_{i=1}^{N} R'_i$, we have

$$E[N'] = E\left[E\left[\sum_{i=1}^{N} R'_i | N\right]\right] \leq E[N] E[K'], \tag{36}$$

which yields

$$E[N] \geq \frac{E[N']}{E[K']}. \tag{37}$$

Since $N \leq n$, we have $E[N] = \Theta(n)$.

By the same argument used in (20) and (21), we can show

$$\text{Var}(N) = O(n \log n). \tag{38}$$

---

[6]Here, in the context of Theorem 10, $b = t_r(n)$, $\sigma^2 = \text{Var}(Z_i) = \text{Var}(\tau) = \Theta(n^2 \log n)$ and $\mu = E[Z_i] = E[\tau] = n$.

Each of these $N$ distinct nodes has at least one version of the given generation for node $j$ with probability $\eta(n)$. Then for any $0 < \epsilon' < 1$, if $\delta n \leq (1-\epsilon')\eta(n)E[N]$,

$$\begin{aligned}
\Pr(\delta n \leq \eta(n)N) &\geq \Pr((1-\epsilon')\eta(n)E[N] \leq \eta(n)N) \\
&= \Pr((1-\epsilon')E[N] \leq N) \\
&\geq 1 - \frac{\text{Var}(N)}{\epsilon'^2 E[N]^2} \\
&= 1 - O\left(\frac{\log n}{n}\right).
\end{aligned} \quad (39)$$

This implies that for any $0 < \epsilon' < 1$, if $\delta \leq \frac{(1-\epsilon')\eta(n)E[N]}{n}$, then node $v$ can collect $m = \delta n$ versions for the given generation within $t_r(n)$ time slots with probability greater than or equal to $1 - O\left(\frac{\log n}{n}\right)$.

By (33), (34) and (37), we have

$$E[N] \geq \frac{E[N']}{E[K']} \geq \frac{E[N']}{\frac{t_r(n)}{n} + 1} = \frac{t_r(n)p_2(n)}{\left(\frac{t_r(n)}{n} + 1\right)}. \quad (40)$$

Therefore, $\frac{E[N]}{n} \geq \frac{p_2 c_1}{c_1 + 1}$ as $n \to \infty$.

By Lemma 5 and Lemma 3, we have $\eta(n) \geq (1 - O(\frac{1}{n}))\phi(n)$ and $\phi(n) \geq (1-\epsilon)\frac{p_s\theta' p_1(n)}{1 + p_s\theta' p_1(n)}$ for any $0 < \epsilon < 1$ with probability $1 - O\left(\frac{\log n}{n}\right)$. Therefore, with probability $1 - O\left(\frac{\log n}{n}\right)$

$$\frac{\eta(n)E[N]}{n} \geq \left(1 - O\left(\frac{1}{n}\right)\right) \frac{p_s\theta' p_1(n)}{1 + p_s\theta' p_1(n)} \cdot \frac{t_r(n)p_2(n)}{t_r(n) + n}. \quad (41)$$

Consequently, as $n \to \infty$, as long as

$$\delta < \frac{p_s\theta_0}{1 + \theta_0 p_s} \cdot \frac{p_2 c_1}{c_1 + 1},$$

then with probability 1, the destination $v$ can collect $m$ versions for the given generation within $\tilde{T}(n) = c_0 \log n + T'_1 + t_r(n) = c_1 n + T'_1 - \log n$ time slots. By the fact that $T(n) \leq_{st} \tilde{T}(n)$ and $T'_1 = o(\log n)$, we have $T(n) \leq c_1 n$ w.h.p. □

We are now ready to prove the main results of this paper.

*Theorem 8:* Let $D_{\pi_{2rsc}}(n)$ be the packet delay of the 2HRRSC scheme under the random walk mobility model. If $m = \delta n$ for $\delta$ satisfying (32), and $c_1 < \frac{1}{\theta' p_s}$, then $D_{\pi_{2rsc}}(n) = \Theta(n)$ w.h.p.

*Proof:* Note that the packet delay is defined as the time from the instant when the packet starts transmission at the source until the instant when the packet is decoded successfully at the destination (Definition 2).

When a source has a generation to send, it transmits (at $n$ time instants) $n$ versions. This takes the source on average $\frac{n}{p_s \theta' p_1(n)}$ time slots. Hence we have $D_{\pi_{2rsc}}(n) = \Omega(n)$.

After these $n$ versions are sent out, they are distributed in the network. All nodes other than the source and the destination act as relays and put the versions into their designated queues for the given S-D pair. Each relay node has a separate queue for that given S-D pair. Note that since the HOL versions in these $n-2$ queues designated for the given S-D pair are not necessarily for the same generation, the destination does not in general collect versions according to the generation order. For instance, suppose the destination has successfully decoded generations $g_1, g_2, ..., g_{i-1}$, and starts to collect versions for $g_i$. During the subsequent collecting process, the destination may collect versions not only for generation $g_i$ but also for generations $g_{i+1}, g_{i+2}, ....$. This is because the relay nodes which have successful transmissions when they meet the destination and choose the destination as the target receiver may not have versions for generation $g_i$. All the same, the relay nodes deliver their HOL versions, for generations $g_j, j \geq i+1$, to the destination. The destination can then decode generation $g_j, j \geq i$, whenever it collects $m$ versions for $g_j$.

Now consider a virtual system in which relay nodes deliver the versions only for the earliest (having the smallest index) undecoded generation to a destination. If a relay node does not have a version for the earliest undecoded generation when it meets the destination, the relay node does not transmit anything to the destination. Hence, in this virtual system, for the above example, the destination will not collect versions for generation $g_j, j \geq i+1$, until $m$ versions for generation $g_i$ have been collected and $g_i$ has been decoded. Clearly, the packet delay in this virtual system is greater than or equal to the packet delay in the actual system.

In the actual system, each relay node has a separate queue for the given S-D pair. In the virtual system, however, it is more convenient to view the $n-2$ queues at the relay nodes for the given S-D pair as a single queue $Q$. In this virtual queueing system, both the arrival and departure processes are with respect to generations. This is illustrated in Fig. 4. The packet delay in the virtual system consists of three delay components: the transmission delay for a generation before its arrival to queue $Q$ (for sending all $n$ versions), the queueing delay in $Q$, and the service time for the generation (for collecting $m$ versions at the destination).

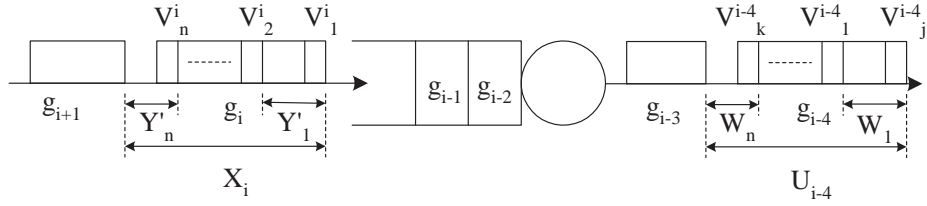

Fig. 4. Virtual queueing system: $g_i$ denotes generation $i$. $V_j^i$ denotes version $j$ of generation $i$. $X_i$ is the inter-arrival time between generation $g_i$ and generation $g_i+1$. $Y_j'$ is time between the transmissions of $V_j$ and $V_{j+1}$. $U_i$ is inter-departure time between $g_i$ and $g_{i+1}$, and $W_j$ is time between the departures of $V_j$ and $V_{j+1}$.

For queue $Q$, the inter-arrival time between generations $g_i$ and $g_{i+1}$ is $X_i$, which is the time between the transmission instants of the first version of $g_i$ and the first version of $g_{i+1}$. Note that $X_i = \sum_{j=1}^{n} Y_j'$, where $Y_j'$ is the time between transmissions of two consecutive versions $V_j^i$ and $V_{j+1}^i$ of generation $g_i$. By (11), we know $E[Y_j'] = \frac{1}{p_s E[\theta(n)]}$. Thus,

$$E[X_i] = \frac{n}{p_s E[\theta(n)]}. \tag{42}$$

Now consider a new queue $Q'$ which has the same potential departure process as $Q$. The inter-arrival time for a generation to $Q'$ is $X_i'' = \sum_{j=1}^{n} Z_j''$, where $Z_j''$ is the time between instants when the source node is in an active cell and chooses to perform as a source. In other words, in the system for $Q'$, the source node sends out a version whenever it is in an active cell and chooses to perform as a source (i.e., it needs not be scheduled). Clearly, the queueing delay of $Q'$ stochastically dominates the queueing delay of $Q$.

Let $Z_k'$ be the time between consecutive returns of the source node to active cells. Under the random walk model, since there are $\theta'n$ active cells with a regular pattern as shown in Figure 1, $Z_k'$ is exactly the return time of a random walk on a $\frac{1}{\sqrt{\theta'}} \times \frac{1}{\sqrt{\theta'}}$ grid. By Lemma 2, we know $E[Z_k'] = \frac{1}{\theta'}$ and $\text{Var}(Z_k') = \Theta(\frac{1}{\theta'^2} \log \frac{1}{\theta'}) = \Theta(1)$. Furthermore, $Z_j'' = Z_1' + Z_2' + \cdots + Z_{H'}'$, where the $Z_k$'s are i.i.d. and $H'$ is a geometric random variable with parameter $p_s$ independent of $Z_k$'s. Then

$$E[Z_j''] = E[H']E[Z_i'] = \frac{1}{p_s \theta'}, \tag{43}$$

and by the law of total variance,

$$\begin{aligned}
\text{Var}(Z''_j) &= \text{Var}(Z'_1 + Z'_2 + \cdots + Z'_{H'}) \\
&= E[\text{Var}(Z'_1 + Z'_2 + \cdots + Z'_{H'}|H')] + \text{Var}(E[Z'_1 + Z'_2 + \cdots + Z'_{H'}|H']) \\
&= E[H']\text{Var}(Z'_k) + \text{Var}(H')E[Z'_k]^2 \\
&= \Theta(1).
\end{aligned} \quad (44)$$

Therefore,

$$E[X''_i] = nE[Z''_j] = \frac{n}{p_s\theta'}, \quad (45)$$

and

$$\text{Var}(X''_i) = n\text{Var}(Z''_j) = \Theta(n). \quad (46)$$

By Lemma 7, we know that after the destination starts to collect versions for a given generation, within $c_1 n$ time slots, w.h.p. the destination can collect enough versions to perform decoding. Consider another queue $Q''$ which has an identical arrival process as $Q'$, but with deterministic inter-service times $U''_i = c_1 n$. Then the queueing delay of $Q''$ is stochastically greater than or equal to that of $Q'$ (as can be shown by coupling methods). By choosing $c_1 < \frac{1}{\theta' p_s}$, it is guaranteed that the generation arrival rate $\frac{\theta' p_s}{n}$ is strictly less than the generation service rate $\frac{1}{c_1 n}$, so that the queue is stable.

By Kingman's bound [17], the queueing delay of $Q''$ is

$$D'' = O\left(\frac{E[X''^2_i] + E[U''^2_i]}{E[X''_i]}\right) = O\left(\frac{\Theta(n^2)}{\Theta(n)}\right) = O(n), \quad (47)$$

which implies $D_{\pi_{2rsc}}(n) = O(n)$. Therefore we have $D_{\pi_{2rsc}}(n) = \Theta(n)$. □

## C. S-D Throughput

We now study the S-D throughput of the proposed 2HRRSC scheme. As the next theorem shows, the 2HRRSC scheme can achieve $\Theta(1)$ throughput under the random walk mobility model. The key point is that if each generation is transmitted $\Theta(n)$ times and the generation has $\Theta(n)$ packets, then each transmission contains $\frac{\Theta(n)}{\Theta(n)} = \Theta(1)$ information. Because every version is transmitted two times (2-hop relay), each S-D pair achieves $\Theta(1)$ throughput.

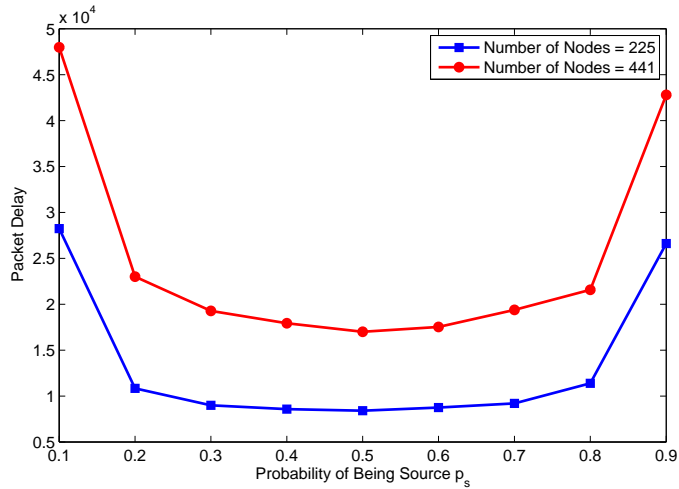

Fig. 5. The effect of probability of being source $p_s$.

*Theorem 9:* Let $T_{\pi_{2rsc}}(n)$ be the S-D throughput of the 2HRRSC scheme under the random walk mobility model. If $m = \delta n$ with $\delta$ satisfying (32) and $c_1 < \frac{1}{\theta' p_s}$, then $T_{\pi_{2rsc}}(n) = \Theta(1)$ w.h.p.

*Proof:* Under the assumptions of the theorem, the proof of Theorem 8 shows that the queueing system corresponding to the network operating under the 2HRRSC scheme has finite $\Theta(n)$ delay. For this stable queueing system, the throughput is the arrival rate. In our setting, the generation arrival rate is $\frac{p_s \theta' p_1(n)}{n}$. The packet arrival rate is $\frac{m p_s \theta' p_1(n)}{n} = \delta p_s \theta' p_1(n)$, which implies that $T_{\pi_{2rsc}}(n) = p_s \theta' p_1(n) \delta$. Since $E[\theta(n)] \to \theta_0$ as $n \to \infty$, we have $T_{\pi_{2rsc}}(n) = \Theta(1)$ w.h.p. □

## V. NUMERICAL EXPERIMENTS

In this section, we present simulation results on the delay performance of the proposed 2HRRSC scheme. In the simulation, we have $n$ nodes uniformly distributed in the unit square, which is divided into $\sqrt{n} \times \sqrt{n}$ cells. The $n$ nodes follow the random walk mobility model as described in Section II. We set $\Delta = \sqrt{2} - 1$ so that

$$\theta' = \frac{1}{(\lceil \sqrt{2}(1+\Delta) \rceil + 1)^2} = \frac{1}{9}.$$

This implies that each center cell of every non-overlapping $3 \times 3$ cell-grid is scheduled as an active cell.

Figure 5 shows the impact of $p_s$, the probability that a scheduled sender chooses to serve as a source, on the packet delay of the 2HRRSC scheme, for the cases of $n = 225$ and $n = 441$, and $\delta = \frac{1}{9}$. As $p_s$ ranges from 0.1 to 0.9, the packet delay performance follows a "U" shape, where the delays are large when $p_s$ is either very small or very large. When $p_s$ is very small, encoded versions are stored in the source for a long period of time, though the versions are delivered from relays to the destination quickly. When $p_s$ is very large, encoded versions are stored in relays' buffers for a much longer period of time, though the versions are sent out from the source more quickly. In the implementation of the 2HRRSC scheme below, we choose $p_s = \frac{1}{2}$.

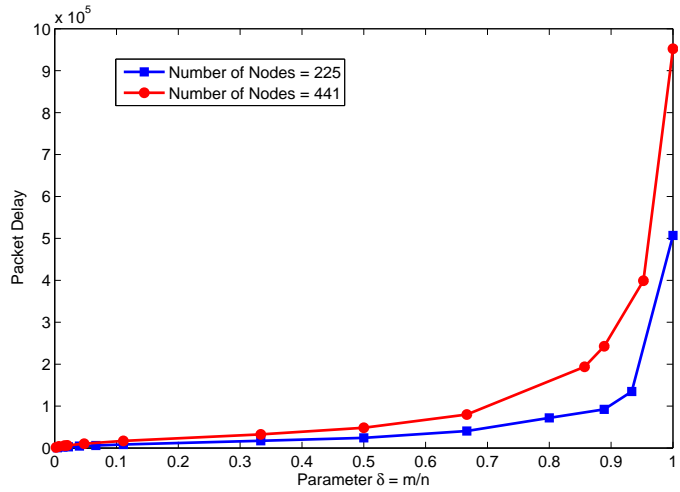

Fig. 6. The effect of parameter $\delta = m/n$.

An important parameter of the proposed 2HRRSC scheme is the encoding ratio $\delta = m/n$, which significantly affects both the throughput and delay performance. Figure 6 shows the effect of $\delta$ on the packet delay of the 2HRRSC scheme with $n = 225$ and $n = 441$ nodes. From the results, we see that when $\delta$ is relatively small (e.g., $\delta < 0.5$), the packet delay increases linearly with $\delta$. On the other hand, when $\delta$ is close to 1 (e.g., $\delta > 0.8$), the packet delay increases exponentially with $\delta$. To understand this behavior, recall that the packet delay consists of three parts: the time required for the source to send out all versions, the queueing delay associated with the relay nodes (cf. proof of Theorem 8), and the time required for the destination to collect enough versions to decode. When $\delta$ is relatively small (e.g., $\delta < 0.5$), all three delay components increase almost linearly with respect to $\delta$. When $\delta$ is close to 1 (e.g., $\delta > 0.8$), although the first delay component grows linearly, the third delay component follows a $\Theta(n \log n)$ growth trend. This is reminiscent of the phenomenon in the coupon collection problem [18]. Also, when $\delta$ becomes large, although the arrival rate of the queueing system of relay nodes (cf. proof of Theorem 8) does not change, the service rate decreases since the destination takes a longer time to collect versions. This causes the second delay component, the queueing delay, to increase exponentially.

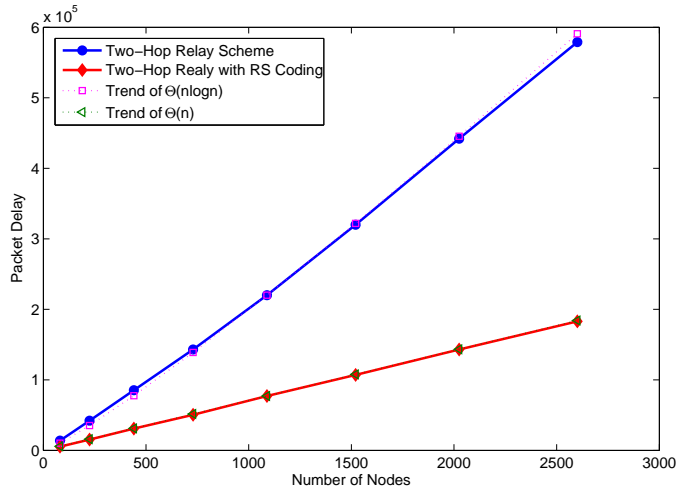

Fig. 7. Packet delay of 2-hop relay scheme and 2HRRSC.

In Lemma 7, we showed that if

$$\delta < \frac{p_s \theta_0}{1 + \theta_0 p_s} \cdot \frac{p_2 c_1}{c_1 + 1},$$

for $2.123 = \frac{4}{e \ln 2} < c_1 < \frac{1}{\theta' p_s} = 18$, then the destination node can collection enough versions for a given generation to decode within $c_1 n$ time slots after the source sends out the $n$th version of the given generation. Note that $0.6798 < \frac{c_1}{c_1+1} < 0.9474$, $p_s = 0.5$, $\theta_0 = \theta'(1 - 2e^{-1}) = \frac{1-2e^{-1}}{9} = 0.0294$, and $p_2 = \theta'(1 - p_s)(1 - 2e^{-1}) = \frac{1-2e^{-1}}{18} = 0.0147$. Hence

$$1.4480 \times 10^{-4} < \frac{p_s \theta_0}{1 + \theta_0 p_s} \cdot \frac{p_2 c_1}{c_1 + 1} < 2.0176 \times 10^{-4}.$$

Therefore, if we choose $\delta \leq 1.4480 \times 10^{-4}$, we can guarantee that the packet delay has a $\Theta(n)$ scaling behavior. Such a small $\delta$, however, is very conservative. In order to show Lemma 7, we have used the conservative bound (32). Choosing a small $\delta$ deteriorates the throughput performance. In real implementations, $\delta$ can be chosen to be much larger than $1.4480 \times 10^{-4}$ or $2.0176 \times 10^{-4}$. For example, Figure 7 compares the packet delay of the 2-hop relay scheme [8] and the 2HRRSC scheme with $p_s = 0.5$, $\theta' = \frac{1}{9}$ and $\delta = \frac{1}{9}$. From the results, we see that the packet delay of the 2HRRSC scheme follows a linear growth with respect to $n$, and also has much smaller value than that of the 2-hop relay scheme [8].

## VI. CONCLUSION

In this paper, we studied the throughput-delay trade-off in mobile wireless networks employing coding techniques. In particular, we proposed a 2-Hop Relay with Reed-Solomon Coding (2HRRSC) scheme, and showed that the proposed scheme can simultaneously achieve $\Theta(1)$ throughput and $\Theta(n)$ delay under the random walk mobility model. The delay improvement from $\Theta(n \log n)$ to $\Theta(n)$ under the random walk type mobility model is significant. This improvement is achieved by combining the diversity introduced by mobility with a reduction in delay introduced by RS coding technique. In contrast to the 2-hop relay scheme in [8], where the destination needs to collect any given packet from a particular relay node, in the 2HRRSC scheme, the destination can collect encoded packets from any of a number of relay nodes it encounters. In a mobile network with $n$ nodes, the 2HRRSC scheme employs an $(m, n)$ Reed Solomon code with an appropriately chosen parameter $m = \Theta(n)$. The scheme guarantees that the destination node can decode the original $m$ packets from the source after collecting any $m$ encoded packets within $\Theta(n)$ time slots. The proposed 2HRRSC scheme not only provides a signicant improvement in the network throughput-delay tradeoff, but also offers potential benets in terms of network robustness and security.

## APPENDIX A

*Theorem 10 (Theorem 5 in [19]):* Let $X_1, X_2, ...$ be i.i.d. with mean $\mu > 0$ and finite positive variance $\sigma^2$. Let $S_0 = 0$, $S_n = \sum_{i=1}^{n} X_i$ and define $\tau(b) = \inf\{n : S_n > b\}$. Then

$$\text{Var}(\tau(b)) = \frac{b\sigma^2}{\mu^3} + \frac{K}{\mu^2} + o(1), \qquad (48)$$

where $K$ is a constant.

## APPENDIX B

*Proof of Lemma 5:* By Lemma 4, we know after $c_0 \log n$ time slots with $c_0 \geq 3/(1 - \frac{1}{n})$, the probability that any given node $u$ is in any given cell $(i, j)$ is bounded between $p_3 = \frac{1}{n} - \frac{1}{n^2}$ and $p_4 = \frac{1}{n} + \frac{1}{n^2}$.

Senders are scheduled uniformly at random in each active cell. Consider an active cell $(i, j)$. $\eta(n)$ is the probability that the sender scheduled in cell $(i, j)$ has a version of generation $g$. Let $k_1$ be the number of nodes within cell $(i, j)$ that have versions of generation $g$, and let $k_2$ be the number of nodes in $(i, j)$ that do not have versions of generation $g$. Then

$$\begin{aligned}
\eta(n) &= \sum_{i=1}^{M} \sum_{j=0}^{n-M} \frac{i}{i+j} \Pr(k_1 = i) \Pr(k_2 = j) \\
&\leq \sum_{i=1}^{M} \sum_{j=0}^{n-M} \frac{i}{i+j} \binom{M}{i} p_4^i (1-p_3)^{M-i} \binom{n-M}{j} p_4^j (1-p_3)^{n-M-j} \\
&= \sum_{i=1}^{M} \sum_{j=0}^{n-M} \frac{i}{i+j} \binom{M}{i} \binom{n-M}{j} p_4^{i+j} (1-p_3)^{n-i-j} \\
&= \sum_{k=1}^{n} \sum_{i=1}^{\min\{k,M\}} \frac{i}{k} \binom{M}{i} \binom{n-M}{k-i} p_4^k (1-p_3)^{n-k} \\
&= \sum_{k=1}^{M} p_4^k (1-p_3)^{n-k} \sum_{i=1}^{k} \frac{i}{k} \binom{M}{i} \binom{n-M}{k-i} + \sum_{k=M+1}^{n} p_4^k (1-p_3)^{n-k} \sum_{i=1}^{M} \frac{i}{k} \binom{M}{i} \binom{n-M}{k-i} \\
&= \sum_{k=1}^{M} p_4^k (1-p_3)^{n-k} \frac{M}{k} \sum_{i=1}^{k} \binom{M-1}{i-1} \binom{n-M}{k-i} + \sum_{k=M+1}^{n} p_4^k (1-p_3)^{n-k} \frac{M}{k} \sum_{i=1}^{M} \binom{M-1}{i-1} \binom{n-M}{k-i} \\
&= \sum_{k=1}^{M} p_4^k (1-p_3)^{n-k} \frac{M}{k} \sum_{i=0}^{k-1} \binom{M-1}{i} \binom{n-M}{k-1-i} + \sum_{k=M+1}^{n} p_4^k (1-p_3)^{n-k} \frac{M}{k} \sum_{i=0}^{M-1} \binom{M-1}{i} \binom{n-M}{k-1-i} \\
&= \sum_{k=1}^{M} p_4^k (1-p_3)^{n-k} \frac{M}{k} \binom{n-1}{k-1} + \sum_{k=M+1}^{n} p_4^k (1-p_3)^{n-k} \frac{M}{k} \binom{n-1}{k-1} \\
&= \sum_{k=1}^{n} p_4^k (1-p_3)^{n-k} \frac{M}{k} \binom{n-1}{k-1} \\
&= \sum_{k=1}^{n} p_4^k (1-p_3)^{n-k} \frac{M}{n} \binom{n}{k} \\
&= \phi(n)[(1+p_4-p_3)^n - (1-p_3)^n] \\
&= \phi(n) \left[ \left(1 + \frac{2}{n^2}\right)^n - \left(1 - \frac{1}{n} + \frac{1}{n^2}\right)^n \right] \\
&= \phi(n) \left(1 + O\left(\frac{1}{n}\right)\right),
\end{aligned}$$

where we used Vandermonde's identities $\sum_{i=0}^{k-1} \binom{M-1}{i} \binom{n-M}{k-1-i} = \binom{n-1}{k-1}$ for $k \leq M$, and $\sum_{i=0}^{M-1} \binom{M-1}{i} \binom{n-M}{k-1-i} = \binom{n-1}{k-1}$ for $k \geq M+1$. These identities can be justified by expanding $(1+x)^{M-1}(1+x)^{n-M} = (1+x)^{n-1}$.

By the same approach, we have

$$\begin{aligned}
\eta(n) &\geq \phi(n)[(1+p_3-p_4)^n - (1-p_4)^n] \\
&= \phi(n) \left[ \left(1 - \frac{2}{n^2}\right)^n - \left(1 - \frac{1}{n} - \frac{1}{n^2}\right)^n \right] \\
&= \phi(n) \left(1 - O\left(\frac{1}{n}\right)\right).
\end{aligned}$$

Therefore, $\eta(n) = \phi(n)$ w.h.p. □